\documentclass[%
 reprint,
 amsmath,amssymb,
 aps,
]{revtex4-1}

\usepackage{graphicx}
\usepackage{dcolumn}
\usepackage{bm}
\usepackage{dsfont}
\usepackage{appendix}

\usepackage{color}
\usepackage{ulem}

\begin{document}

\preprint{APS/123-QED}

\title{A solution for the doublet-triplet splitting in $SU(5)$ }

\author{Renata Jora
	$^{\it \bf a}$~\footnote[1]{Email:
		rjora@theory.nipne.ro}}

\affiliation{$^{\bf \it a}$ National Institute of Physics and Nuclear Engineering PO Box MG-6, Bucharest-Magurele, Romania}

\begin{abstract}
We present a solution to the doublet-triplet splitting problem in supersymmetric $SU(5)$ by enlarging the group to $SU(5)\times SU(5)$ and by introducing two supermultiplets in the $(5,5)$ and $(\bar{5},\bar{5})$ representations of the extended group. The non-perturbative superpotential associated to this gauge group provides not only the correct gauge symmetry breaking but also  the large mass for the Higgs triplets associated to massless Higgs doublet. The method has all the advantages of the sliding singlet model but without the drawbacks.

\end{abstract}
\maketitle

One of the best alternatives for extending the standard model of elementary particles in the ultraviolet regime  is the gauge coupling unification under a single gauge group or product of groups $SU(5)$, $SO(10)$, $SU(4)\times U(1)$, $SU(5)\times SU(5)$ or $E_6$ \cite{Georgi1}-\cite{Langacker}. Among these the simplest and most amenable framework is $SU(5)$ where all standard model matter can be fit into three families of the antifundamental and antisymmetric representations. A step forward into the naturalness direction would then be to consider a supersymmetric version of $SU(5)$ \cite{sup} considered as a natural extension of MSSM.

In supersymmetric $SU(5)$ the three families of matter are arranged in supermultiplets \cite{Mohapatra} in the $\bar{5}$,
\begin{eqnarray}
\bar{F}=
\left(
\begin{array}{c}
d_1^c\\
d_2^c\\
d_3^c\\
e^{-1}\\
\nu_e
\end{array}
\right),
\label{firstmult76}
\end{eqnarray}
and $10$ dimensional representations:
\begin{eqnarray}
T(10)=
\left(
\begin{array}{ccccc}
0&u_3^c&-u_2^c&u_1&d_1\\
-u_3^c&0&u_1^c&u_2&d_2\\
u_2^c&-u_1^c&0&u_3&d_3\\
-u_1&-u_2&-u_3&0&e^+\\
-d_1&-d_2&-d_3&-e^+&0
\end{array}
\right).
\label{secmult76756}
\end{eqnarray}

The supersymmetric $SU(5)$ extensions of the Higgs multiplets are situated in the fundamental $H(5)$ and antifundamental $\bar{H}(\bar{5})$ of the model. In general additional matter multiplets are necessary in order to break the GUT gauge group down to $SU(3)_c\times SU(2)_L\times U(1)_Y$. The minimal choice is  the addition of a  matter multiplet $\Sigma$ in the adjoint representation of the gauge group. The superpotential is  given by:
\begin{eqnarray}
&&W=\frac{1}{2}fV{\rm Tr}(\Sigma^2)+\frac{1}{3}f{\rm Tr}(\Sigma^3)+
\nonumber\\
&&\lambda\bar{H}_{\alpha}\Sigma^{\alpha}_{\beta}H^{\beta}+m\bar{H}H+
\nonumber\\
&&\sqrt{2}y_d^{ij}T_i^{\alpha\beta}\bar{F}_{j\alpha}\bar{H}_{\beta}+\frac{1}{4}Y_u^{ij}\epsilon^{\alpha\beta\gamma\delta\sigma}T_i^{\alpha\beta}T_j^{\gamma\delta}H^{\sigma}.
\label{supepot768856}
\end{eqnarray}

The desired gauge symmetry breaking is obtained for a vacuum expectation value:
\begin{eqnarray}
\langle \Sigma\rangle=
V\left(
\begin{array}{ccccc}
2&0&0&0&0\\
0&2&0&0&0\\
0&0&2&0&0\\
0&0&0&-3&0\\
0&0&0&0&-3
\end{array}
\right).
\label{symbr657}
\end{eqnarray}

Upon gauge symmetry breaking the Higgs supermultiplets may be written as:
\begin{eqnarray}
H=
\left(
\begin{array}{c}
\xi_u\\
H_u
\end{array}
\right)
\,\,\,\,
\bar{H}=
\left(
\begin{array}{c}
\bar{\xi}_d\\
H_d
\end{array}
\right),
\label{higgsmult76}
\end{eqnarray}
where $\bar{\xi}_d$ and $\xi_u$ are the Higgs triplets and $H_u$ and $H_d$ are the Higgs doublets.
The Higgs multiplets have no vacuum expectation values at the GUT scale.
The effective part of the superpotential that contains the mass terms for the Higgs multiplets is:
\begin{eqnarray}
W_m=\lambda\bar{H}_{\alpha}\langle \Sigma^{\alpha}_{\beta}\rangle H^{\beta}+m\bar{H}H.
\label{effsuptpot75664}
\end{eqnarray}

In order for the triplets $\xi_u$ and $\bar{\xi}_d$ to be absent from the low energy theory and to obtain a massless Higgs doublet one would then require that:
\begin{eqnarray}
m=3V\delta^{\alpha}_{\beta},
\label{massterm4554}
\end{eqnarray}
which means that the desired cancellation must be introduced by hand. This is the known doublet-triplet splitting problem \cite{Mohapatra}. A few  solutions have been proposed to regulate this issue. The first one, the sliding singlet model \cite{Witten} involves the introduction of an additional singlet $Z$ that couples with the Higgs multiplets such that the superpotential becomes:
\begin{eqnarray}
W_1=\bar{H}\Sigma H+Z\bar{H}{H}.
\label{supeprt5664}
\end{eqnarray}
In this case the minimum equation for $H$ leads to:
\begin{eqnarray}
-3V+\langle Z\rangle=0,
\label{sol7688}
\end{eqnarray}
as wanted. However after supersymmetry breaking one loop corrections to $\langle Z\rangle$  produce GUT scale masses for the Higgs doublet thus spoiling the gauge hierarchy. 

A second solution, the missing partner mechanism \cite{Georgi3}  would imply the introduction of matter supermultiplets in higher representation of the gauge group and also placing $\Sigma $ in the $75$ of $SU(5)$. This complicates the theory but represents an acceptable solution. We just mention here the other two proposals for resolving this problem, the Dimopoulos-Wilczek mechanism \cite{Dimopoulos} and the GIFT mechanism \cite{Kakuto}. An updated version for the sliding singlet mechanism which solves some of the issues was introduced later in \cite{Barr} (see also the references there).

In this work we will propose a solution to the doublet-triplet problem that has the advantages of the sliding singlet model without the drawbacks. We extend the standard supersymmetric $SU(5)$  to the enlarged group $SU(5)\times SU(5)$ where all matter multiplets present in the first $SU(5)$ are singlets under the second $SU(5)$ and we introduce the additional matter multiplets $\bar{K}$ with the assignment under $SU(5)\times SU(5)$ $(\bar{5}, \bar{5})$ and $K$ with the assignment $(5,5)$. Then at one loop the first $SU(5)$ with the coupling $g_5$ will have the beta function:
\begin{eqnarray}
\beta(g_5)=-\Bigg[3N-5-3(\frac{3}{2}+\frac{1}{2})-1\Bigg]g_5^3=-3g_5^3,
\label{firstbetafucn6455377}
\end{eqnarray}
whereas the second $SU(5)$ with the coupling constant $g_5'$ has:
\begin{eqnarray}
\beta(g_5')=-\Bigg[3N-5\Bigg](g_5')^3=-10(g_5')^3.
\label{secondbteaufnc64553663}
\end{eqnarray}

One notices that the second $SU(5)$ has $N=N_f$ and corresponds exactly to the non-perturbative case discussed for a single gauge group in \cite{Seiberg1},\cite{Seiberg2}. One can introduce the meson fields $M^A_B=\sum_a \bar{K}_{aA}K^{aB}$ and the baryons $B$ and $\bar{B}$ where the small letter indices refer to the second $SU(5)$ whereas the capital letter indices refer to the first $SU(5)$. It is then natural to consider that the relevant part of the new superpotential that contains fields from the second $SU(5)$ is given by:
\begin{eqnarray}
W_m=a(\det M+\bar{B}B-\Lambda_2^{10})+b\bar{H}_AM^A_BH^B.
\label{newsupepot65774}
\end{eqnarray}
Here $\Lambda_2$ is the non-perturbative scale associated with the second group $SU(5)$. For completeness we denote the non-perturbative scale associated with the first group $SU(5)$, $\Lambda_1$. 

The supersymmetric minimum for $M$ is:
\begin{eqnarray}
\det(M)\frac{1}{M^A_B}=0.
\label{rez54663}
\end{eqnarray}
First we notice that $M^A_B=S\delta^A_B+Q^A_B$ where $S$ is the first $SU(5)$ singlet and $Q$ is a traceless matrix corresponding to the adjoint $24$ representation of the first $SU(5)$. The first solution is then $M^A_B=0$ which does not break the first $SU(5)$. The second choice is two diagonal values of zero and then the rest of three equal and derived from the trace condition. We denote:
\begin{eqnarray}
&&\langle S\rangle=s
\nonumber\\
&&\langle M^A_A(A=1..3)\rangle=s+x
\nonumber\\
&&\langle M^A_A(A=4,5)\rangle=s+y
\label{notyr6577}
\end{eqnarray}
Then one can choose $y+s=0$ and from $3x+2y=0$, $x=-\frac{2}{3}y$. The meson $Q^A_B$ has the correct vacuum expectation values to break the first group $SU(5)$  down to the standard model group. Normally the singlet $S$ does not couple with the gauge group. It is then evident from Eq. (\ref{newsupepot65774}) that the mass term for the triplet $\bar{\xi}_d$ and $\xi_u$ are at the GUT scale whereas $H_u$ and $H_d$ remain massless. Moreover the determinant term induces GUT scale masses for $S$ and $Q^A_B$ with $A,B=4,5$ according to:
\begin{eqnarray}
&&a\det(M)=a\Bigg[(-\tilde{M}^4_5\tilde{M}^5_4+\tilde{M}^4_4\tilde{M}^5_5)(s+x)^3+
\nonumber\\
&&{\rm higher\,\,order\,\,interaction\,\,terms}\Bigg].
\label{mass6455}
\end{eqnarray}
Here $M=\langle M\rangle +\tilde{M}$.

 When supersymmetry is broken one loop corrections to $Q^C_D$ with $D.C=1..3$ couple only to Higgs triplet thus preserving the desired gauge hierarchy. The model has all the advantages of the sliding singlet model without the disadvantages. The fields $\Sigma$ and $S$ then gain masses at the GUT scale by integrating out the heavy Higgses. The second $SU(5)$ group is broken completely with masses at the corresponding gauge bosons at the GUT scale.

We assume that at the Planck scale the two $SU(5)$ group unify under a single gauge group with a single gauge group. Knowing that $M_G=2\times 10^{16}$ GeV where $M_G$ is the GUT scale and $\frac{1}{\alpha_G}=24.3$ where $\alpha_G$ is the first $SU(5)$ group coupling constant at the GUT scale \cite{PDG} from,
\begin{eqnarray}
\frac{1}{\alpha_G}=\frac{3}{2\pi}\ln(\frac{M_G}{\Lambda_1}),
\label{integ5466355}
\end{eqnarray}
one can determine $\Lambda_1=1.578\times 10^{-6}$ GeV. Furthermore the condition of unification at the Planck scale leads to:
\begin{eqnarray}
\frac{3}{8\pi^2}\ln(\frac{m_P}{\Lambda_1})=\frac{10}{8\pi^2}\ln(\frac{m_P}{\Lambda_2}),
\label{unifc6577}
\end{eqnarray}
which yields $\Lambda_2=4.12\times 10^{11}$ GeV and $g_P=0.678$ where $g_P$ is the unification coupling constant at the Planck scale.

In summary we presented a straightforward solution for the doublet-triple splitting problem that requires no fine tuning of the terms in the superpotential and has no significant drawbacks.

\end{document}